\documentclass[12pt]{iopart}

\usepackage[utf8]{inputenc}
\usepackage{graphicx}
\usepackage{amssymb}
\usepackage{hyperref}

\newcommand{\cns}{CE$\nu$NS}
\newcommand{\weakangle}{\sin^2\theta_{W}}
\newcommand{\isotope}[2]{$^{#2}{\rm #1}$}

\begin{document}

\title[Ricochet at Chooz]{Coherent Neutrino Scattering with Low Temperature Bolometers at Chooz Reactor Complex}

\author{J. Billard$^1$, R.~Carr$^2$, J. Dawson$^3$, E. Figueroa-Feliciano$^4$, J. A. Formaggio$^2$, J. Gascon$^1$, , S.~T.~Heine$^{2}$, M. De Jesus$^1$, J. Johnston$^2$, T. Lasserre$^{3,5}$, A. Leder$^2$, K. J. Palladino$^6$, V.~Sibille$^{2}$, M. Vivier$^5$, and L. Winslow$^2$}

\address{\it $^1$ Univ Lyon, Université Claude Bernard Lyon 1, CNRS/IN2P3, Institut de Physique nucléaire de Lyon, 4 rue Enrico Fermi, F-69622 France}
\address{\it $^2$ Laboratory for Nuclear Science, Massachusetts Institute of Technology, Cambridge, MA, USA}
\address{\it $^3$ APC, AstroParticule et Cosmologie, Universit\'e Paris Diderot, CNRS/IN2P3, CEA/Irfu, Observatoire de Paris, Sorbonne Paris Cité, 10, rue Alice Domon et Léonie Duquet, 75205 Paris Cedex 13, France}
\address{\it $^4$ Department of Physics, Northwestern University, Evanston, IL, USA}
\address{\it $^5$ Commissariat \`{a} l'\'energie atomique et aux \'energies alternatives, Centre de Saclay, DRF/Irfu, 91191 Gif-sur-Yvette, France}
%
%
\address{\it $^6$ Department of Physics, University of Wisconsin, Madison, WI, USA}
\ead{lwinslow@mit.edu}
\vspace{10pt}
\begin{indented}
\item[]\today
\end{indented}

\begin{abstract}
We present the potential sensitivity of a future recoil detector for a first detection of the process of coherent elastic neutrino nucleus scattering (\cns).  We use the Chooz reactor complex in France as our luminous source of reactor neutrinos.  Leveraging the ability to cleanly separate the rate correlated with the reactor thermal power against  (uncorrelated) backgrounds, we show that a 10 kilogram cryogenic bolometric array with 100 eV threshold should be able to extract a \cns~signal within one year of running.
\end{abstract}
\vspace{2pc}
\vspace{2pc}
\submitto{\jpg}
\maketitle

\section{Introduction}
It has been almost four decades since the idea that neutrinos should be able to coherently scatter against the nucleus as a whole was originally proposed~\cite{Freedman:1973yd}. The condition for such coherence is typically met for neutrinos with energies $\lesssim$100~MeV. In this energy range, coherent elastic neutrino nucleus scattering (\cns) possesses a large cross-section on neutron rich targets when compared to other common detection channels such as inverse beta decay and neutrino electron scattering. Despite this clear enhancement, the \cns~process has never been observed due to the difficulty in detecting the low-energy $\sim$0.1-10~keV nuclear recoil that marks its signature. A first detection of this process would be an important test of the Standard Model. Subsequent measurements could search for non-standard neutrino interactions~\cite{scholberg2006} and sterile neutrinos~\cite{ricochet1}. The process also makes an excellent probe of the poorly constrained neutron density functions that dominate the calculation of the nuclear form function~\cite{Patton:2012jr,McLaughlin:2015xfa}, which is especially critical to understanding the production and detection of supernova neutrinos~\cite{wilsonSN,horowitzSN,scholbergSN,Horowitz:2003cz}. The physics implications are broad, from the physics of neutron stars to the neutrino floor for direct dark matter experiments~\cite{Billard:2013qya}. Finally, observation of \cns~ has potential practical applications, since the process could also be used as a tool for nuclear reactor monitoring~\cite{Christensen:2014pva}.

In order to meet the energy requirement for coherence, along with basic detection constraints of a sufficiently high flux and nuclear recoil energy, there are three possible sources of (anti)neutrinos that could be used to search for \cns: pion decay at rest beam (DAR), an intense radioactive source, and a nuclear reactor. The DAR source has the highest energy neutrinos and therefore the energy threshold needed to detect \cns~has already been achieved in germanium, sodium and liquid argon-based detectors.  The COHERENT experiment proposes to use the Spallation Neutron source as a DAR and a suite of two if not three of the above detector technologies~\cite{coherent}. TEXONO has a similar proposal for Ge-based detectors~\cite{texano}. 

The program becomes more challenging if one is to detect \cns~ from a nuclear reactor or an intense radioactive source.  The mean neutrino energy at a nuclear reactor such as Chooz is around 3.6 MeV with an endpoint near 10 MeV.  Due to the lower neutrino energies, recoil detector technologies with sub-keV energy thresholds are required. cryogenic bolometers could be optimized to achieve such low energy thresholds, as advanced by the Ricochet experiment~\cite{ricochet1}. The CONNIE experiment proposes to achieve the needed threshold using charge coupled devices (CCDs)~\cite{connie,Aguilar-Arevalo:2016qen}. As will be discussed below, metallic superconducting bolometers and Ge-based semiconductors can be used to achieve these low thresholds.

Maximizing the advantage provided by low threshold detectors and the high flux of commercial reactors could provide the ideal combination to make a definitive measurement of the \cns~process.  France's Chooz Nuclear Power Plant is a uniquely well-suited location for a reactor-based \cns~search. Each of the site's twin reactors generates 4.3 GW of thermal power during standard operations, making them among the most luminous reactor neutrino sources in the world~\cite{Ardellier:2006mn}. The high neutrino flux makes it feasible to position a detector well outside the reactor containment buildings, eliminating the reactor-produced neutron and gamma backgrounds that challenge close-to-core measurements. At Chooz, suitable laboratories already exist currently used by the Double Chooz near detector with a 140 m.w.e. overburden reduces the cosmic ray flux down to $(3.74 \pm 0.21)~\mu$/m$^{2}/$s~\cite{Dietrich:2013}. Comparing data collected when both reactors are operating to data collected during single-reactor refueling shutdowns will provide both a powerful background discriminant and a demonstration of \cns~as a reactor monitoring tool.

We examine the sensitivity of a future bolometer-based experiment, Ricochet, located at the near site location of the Chooz Nuclear Power Plant.  We consider a composite detector of both Ge-based and metallic Zn-based smaller detectors with a maximum total mass of 10 kg.  We show that, even assuming minimal knowledge of background rates, we can measure the \cns~rate in the proposed detector to a precision of 35\% or better after one year of running.

\section{Low Threshold Cryogenic Detectors}

-

Leveraging these two different background rejection techniques within the same detector would give greater confidence for determining the detection of the~\cns~process. However, for this paper, we do not explicitly assume any specific background rejection technique, but rather consider the potential of identifying a signal correlated with reactor power against an uncorrelated background. 

\section{Neutrino Signal from Chooz}

The Chooz reactor complex in Chooz, France provides an optimal setting for a first detection of the \cns~process.  The complex comprises two commercial nuclear reactors of the same design with a combined thermal power of 8.54 GW.  We imagine placement of a combined 10-kg payload at the site of the Double Chooz experiment's near detector, which is located 355.39 m and 468.76 m away from the two reactor cores.  For the calculation of the flux, we assume the neutrino spectrum as calculated by~\cite{PhysRevC.84.024617} for \isotope{U}{235},\isotope{Pu}{239}, and \isotope{Pu}{241}, and \cite{PhysRevC.83.054615} to calculate the \isotope{U}{238} spectrum.  For this study, we assume the fraction of the four isotopes are (\isotope{U}{235}:\isotope{Pu}{239}:\isotope{Pu}{241}:\isotope{U}{238} = 55.6\%:32.6\%:7.1\%:4.7\%)\cite{Ardellier:2004ui}.  Such particular combination is representative of typical conditions of the Chooz reactor. Changes in composition over one reactor cycle are ignored for this study becuase they cause the rates to deviate by less than 2.5\%. We approximate the neutrino energy spectrum as constant below 2 MeV, since the phenomenological approximation used to determine the neutrino flux are typically not reported below this energy.

The \cns~ cross-section for a target nucleus $(A)$ as a function of its recoil energy $E_R$ is given by the expression:

\begin{equation}\label{eq:cross}
\frac{d\sigma(\nu A \rightarrow \nu A)}{dE_R}  = \frac{G_F^2}{4\pi} M_A Q_W^2 (1 - \frac{M_A E_R}{2 E_\nu^2}) F(q^2)^2
\end{equation}
\noindent where $G_F$ is the Fermi coupling constant, $M_A$ is the mass of the nucleus, $F(q^2)$ is the nuclear form factor~\cite{McLaughlin:2015xfa},  $Q_W= N - Z(1-4\weakangle)$ is the weak charge, and $N(Z)$ represents the number of neutrons (protons) in the nucleus.  Due to the Weinberg angle suppression, the process  scales primarily with the number of neutrons, squared.

\begin{figure}[htb]
\begin{center}
\includegraphics[height=3in,width=4in]{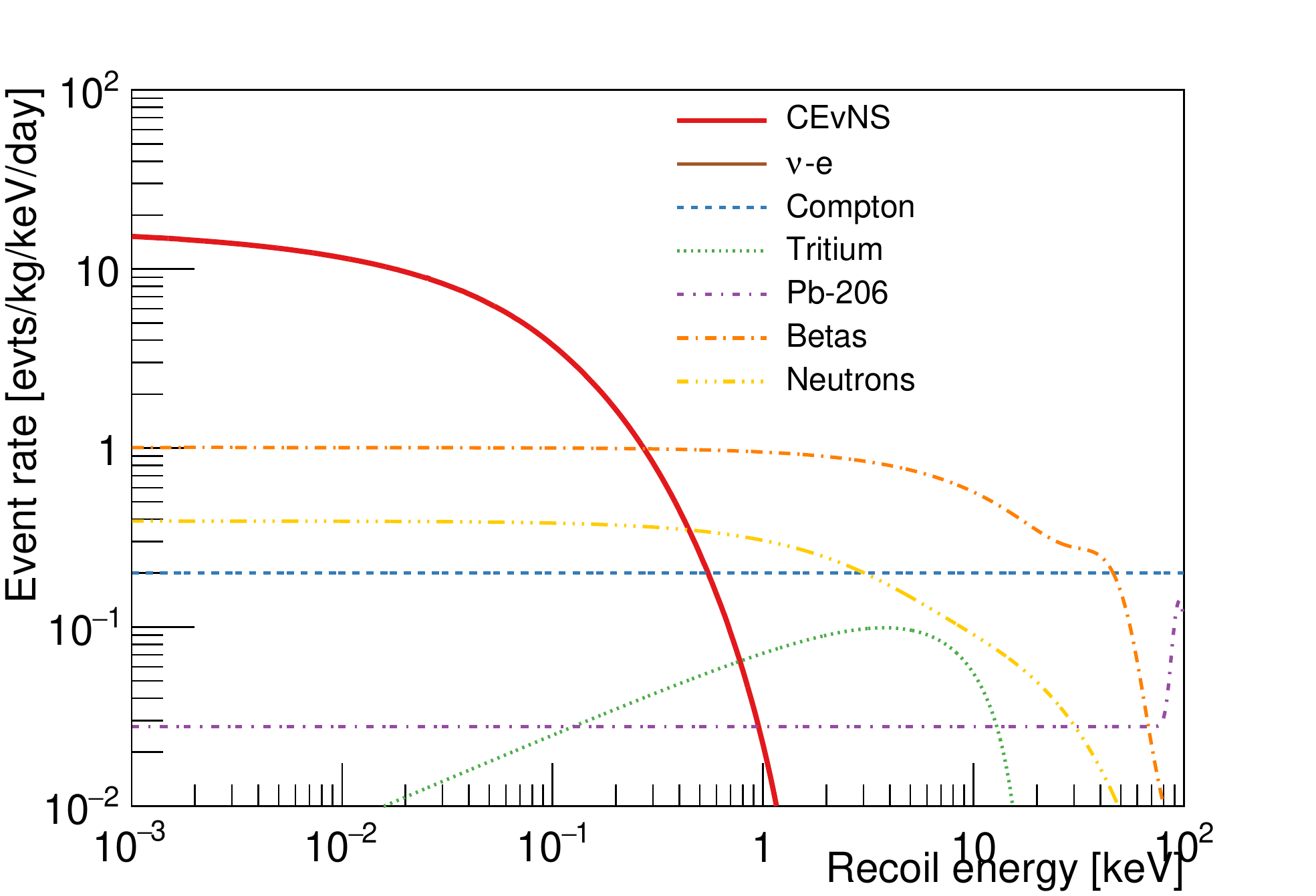}
\caption{The differential \cns~ rate and projected background rates versus detector recoil energy. The internal backgrounds, such as tritium and \isotope{Pb}{206}, are based on observed rates as measured by the EDELWEISS-III experiment \cite{Hehn:2016nll}}.
\label{fig:diff_rate}
\end{center}
\end{figure}

Figure \ref{fig:diff_rate} shows the differential rate for the predicted neutrino signal versus detected recoil energy. By integrating above a certain threshold, we can determine the total event rates (Table \ref{tab:diff_rate}).  For a 10-kg payload equally divided into Ge- and Zn based detectors, we would expect $\sim 4.85$ events/day with a 100 eV recoil threshold and full reactor power.  This rate increases by almost 50\% if the threshold drops to 50 eV, placing a high premium on the detector threshold one can achieve.

\begin{table}
\begin{center}
\begin{tabular}{ |c|cccc| }
\hline
Detector & \multicolumn{4}{c|}{Rate (per kg per day)} \\
Threshold & Si & Zn & Ge & Os \\
\hline
50 eV & 0.32 & 0.66 & 0.76 & 1.25 \\
\hline
100 eV & 0.26 & 0.46 & 0.51 & 0.55 \\
\hline
200 eV & 0.19 & 0.25 & 0.26 & 0.14 \\
\hline
\end{tabular}
\caption{The predicted \cns~event rate for various detector materials and recoil threshold energies. Rates are calculated for a reactor power of 8.54 GW, a detector distance of 400 m, and fission fractions \isotope{U}{235} = 55.6\%, \isotope{Pu}{239}= 32.6\%,\isotope{Pu}{241} = 7.1\%, and \isotope{U}{238} = 4.7\%.  We assume an overall $\pm 5\%$ uncertainty on the predicted rate of neutrino interactions.}
\label{tab:diff_rate}
\end{center}
\end{table}

\section{Background Estimates}

Placement of the Ricochet experiment at the Chooz near reactor site --over 400 meters of rock and soil separating the detector from the reactor cores-- has the unique advantage in that it reduces the reactor-correlated neutron and gamma backgrounds to undetectable levels.  This allows for a clean separation of ``neutrino-induced events'', which are correlated with the reactor thermal power, and ``background events'', 5 are uncorrelated with the reactor activity.  Such leverage has already been used by the Double Chooz collaboration with great success in its measurement of the neutrino-induced inverse beta decay process for the purposes of extracting the neutrino mixing angle $\theta_{13}$~\cite{Abe:2014bwa,Abe:2012ar}.  To help determine Ricochet's sensitivity to the \cns~reaction, we will first provide an estimate of the uncorrelated background, as shown below.

\subsection{External Neutrons}
\begin{figure}[htb]
\begin{center}
\includegraphics[width=0.85\textwidth]{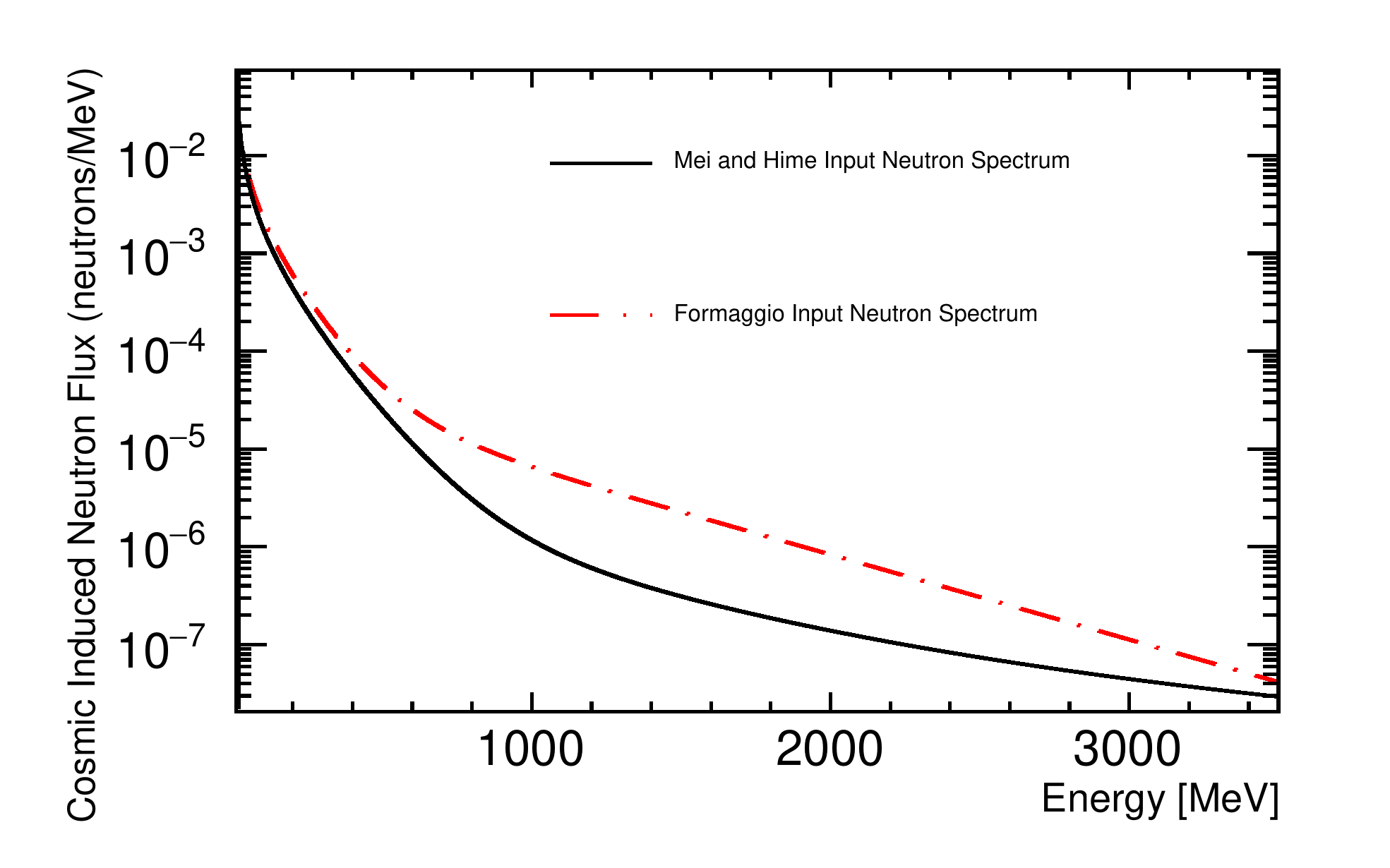}
\caption{The estimated cosmogenic-activated neutron flux as a function of initial neutron energy at the near hall of the Chooz reactor complex. The neutron energy spectra are derived from parametrizations from Mei and Hime~\cite{Mei2006} (solid line) and Formaggio~\cite{Formaggio:2004ge} (dashed line).  Both spectra are normalized to unity.}
\label{ricochet_FN}
\end{center}
\end{figure}

The $(\alpha, n)$ background represents the neutrons created from the naturally occurring uranium and thorium decay chains interacting with the various elements in the surrounding cavern.  These reactions tend to produce neutrons with kinetic energies below 10 MeV.  Combining existing measurements of the rock composition at the Double Chooz far site~\cite{Tang2006} and simulations of the U/Th decay spectra (using GEANT4~\cite{Agostinelli2003250}), we are able to model the neutron spectrum from $(\alpha,n)$ reactions, which is expected to be below $10^{-7}$ events/eV/kg/day, well below other contributions.  

A similar procedure is used to assess the neutron background due to cosmogenic activity surrounding the cavern of the Double Chooz near site.  Neutrons from cosmogenic activation tend to possess much higher energies (limit closer to $\le 100$ GeV), and pose a more serious background source (see Figure~\ref{ricochet_FN}).  The initial neutron energy spectrum is modelled to follow the parametrization as reported in ~\cite{Formaggio:2004ge}, assuming a generating muon energy of 32.1 GeV. The cosmogenic neutron spectra are propagated through GEANT4-based simulations of both the Double Chooz near detector~\cite{JunqueiradeCastroBezerra2015} and the Ricochet-style detector. The number of events observed for the Double Chooz simulation is normalised to the fast neutron rate as measured by the Double Chooz near detector, thereby providing the equivalent run time of the simulation. Comprehensive analyses utilising Gd-captures only~\cite{Ishitsuka2016} or Gd and H-captures~\cite{Cabrera2016} have been performed; the most pessimistic scenario was retained.  For the Ricochet-style detector, we assume a cylindrical water shield with 3.25 meters of water buffering the detector cryostat on its side, and 2 meters of water shielding from the bottom.  The estimated background rate from fast neutrons is $(2.3 \pm 2.3)\times 10^{-4}$ neutrons/kg/eV/day, leading to a total rate of $\left(0.3 \pm 0.1\right)$ neutrons/kg/day for recoil energy depositions between 0.1 and 1 keV.   An additional check using the parametrization of  Mei and Hime~\cite{Mei2006} produced a 95 percent confidence level of a neutron background of less than 0.4 neutrons/kg/day in the same region of interest.

\subsection{Internal Radioactivity}

\begin{figure}[htb]
\begin{center}
\includegraphics[width=0.55\textwidth]{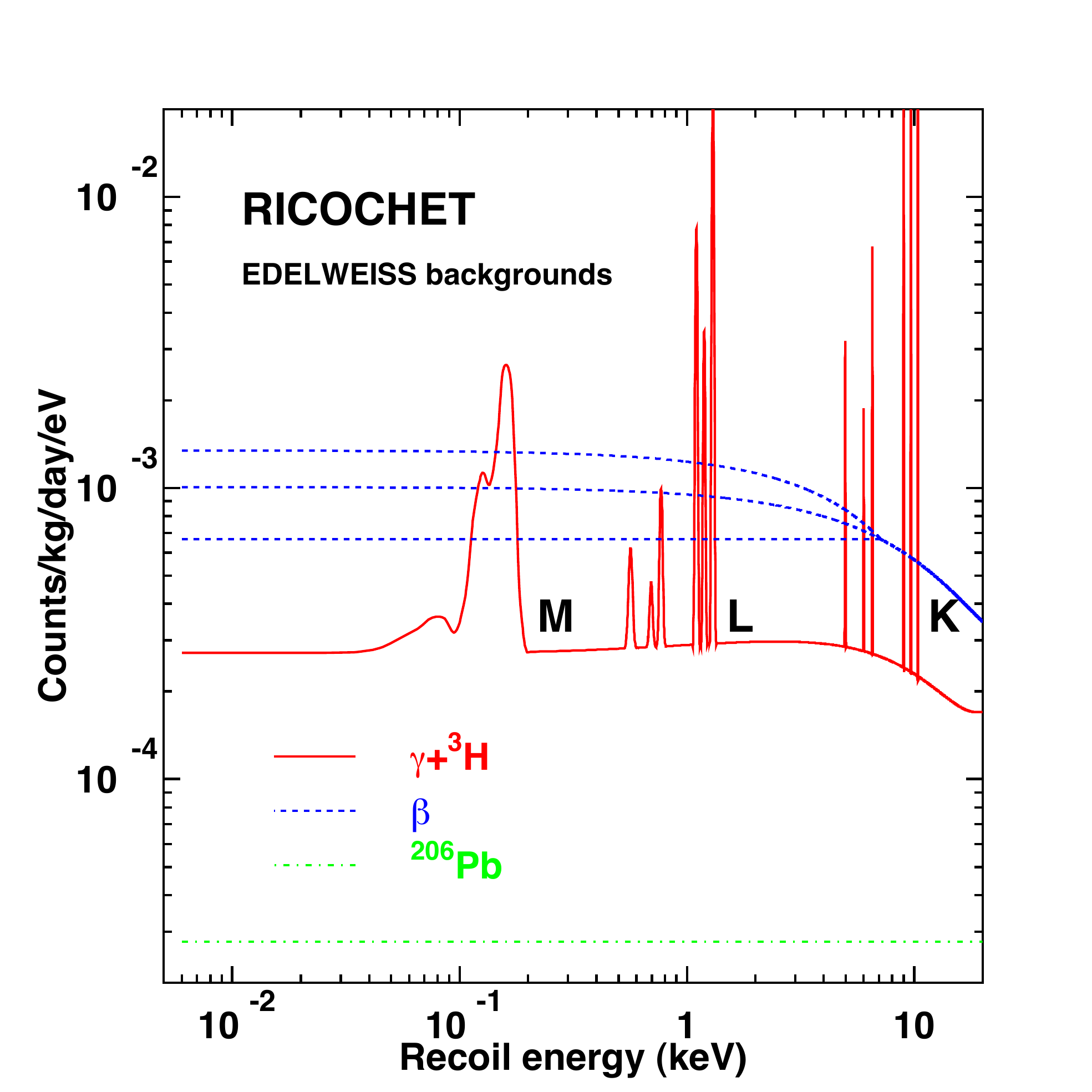}
\caption{The differential event rate of the internal backgrounds of a Ge detector as observed by the EDELWEISS-III experiment \cite{Hehn:2016nll}: gamma + tritium (red solid line), beta (blue dotted line) and \isotope{Pb}{206}(green dot-dashed line), in recoil energy. The different beta lines indicate the spread in uncertainty due to the extrapolation of the background to lower energies.}
\label{ricochet_int_back}
\end{center}
\end{figure}

The internal backgrounds for the Ricochet experiment will be assumed to be similar in magnitude and shape to the ones observed by the EDELWEISS-III dark matter direct detection experiment~\cite{Hehn:2016nll} as the material selection will require the same level of radiopurity guaranteed by screening \cite{scorza}. Figure~\ref{ricochet_int_back} presents the low energy contributions of the different internal backgrounds as observed by the EDELWEISS-III experiment which will be used hereafter for the Ricochet sensitivity estimates. The gamma background (red solid line) is explained by the Compton component from radiogenic gammas and the cosmogenic activation of the Ge crystal leading to EC-lines (10 keV K-shell, 1.3 keV L-shell, 160 eV M-shell) and tritium beta decay with an endpoint of 18.6 keV \cite{Armengaud:2016aoz}. These backgrounds can be kept at a minimal level by maximally reducing the exposure of the detectors to the surface to prevent cosmogenic activation. Surface contamination from \isotope{Rn}{222} leads to the \isotope{Pb}{210} decay chain, characterized by a 22.3 years of half-life, that undergo beta decays (blue dotted line), alphas and \isotope{Pb}{206} nuclear recoils  (green dashed line). The shape and normalization of these spectra are taken from the measured distributions of tagged
surface events in EDELWEISS~\cite{Hehn:2016nll}, which are entirely consistent with a population of \isotope{Pb}{210} in
equilibrium with its progenies. These backgrounds require working under radon-free air combined with efficient chemical etching of copper~\cite{etching}. Radiogenic neutrons are not shown here as they are negligible compared to cosmogenic neutrons as discussed above~\cite{Hehn:2016nll}.

The one long-lived contribution that would not be present in the EDELWEISS germanium detectors that is difficult to extrapolate directly to the metallic zinc detectors is the cosmogenic activation of \isotope{Zn}{65} ($\tau_{1/2} = 243$ days), which has both a $\beta$-decay channel as well as a low energy meta-stable state (\isotope{Zn}{65m}, with emission at 53 keV).  The cosmogenic activation of \isotope{Zn}{65} is not included in our background estimates.

Of note, the isotopes of \isotope{Zn}{64}, \isotope{Zn}{70}, and \isotope{Ge}{76} undergo $\beta\beta$-decay, but due to the large lower bounds on their half lives calculated in \cite{BELLI2009256,BARABASH2015416} and large endpoint energies on the order of 1 MeV, we will detect less than $0.01$ events/kg/day in our region of interest.  A summary of all considered backgrounds can be found in Table~\ref{tab:backgrounds}.

\begin{table}[htp]
\caption{A summary of correlated and uncorrelated backgrounds to reactor power on a 5 kg Ge + 5 kg Zn recoil detector located at the Chooz near site underground hall. The region of interest (ROI) is defined between 100 eV and 1 keV. The tritium estimate($^{\dagger}$) on differential rate includes only beta-spectrum contribution.}
\begin{center}
\begin{tabular}{|l|c|c|}
\hline
Background Source & Rate  & Total Rate in ROI \\
& (kg$^{-1}$day$^{-1}$eV$^{-1}$) & (events/day) \\
\hline
\multicolumn{3}{|l|}{{\em Uncorrelated Rate}} \\
\hline
Gammas (Compton) & $(2.0 \pm 0.2)\times 10^{-4}$ & $1.8 \pm 0.18$ \\
Fast Neutrons & $(3.3 \pm 1.1)\times 10^{-4}$ & $3.0 \pm 1.0$ \\
$(\alpha, n)$ & $ < 1.1 \times 10^{-5}$& $< 0.1$  \\
$\beta$ decays from \isotope{Pb}{210} & $(9.3 \pm 3) \times 10^{-4}$ &  $8.8 \pm 2.8$ \\
\isotope{H}{3}$^{\dagger}$ & $(0.36 \pm 0.04) \times 10^{-4}$ &  $0.5 \pm 0.05$ \\
\isotope{Pb}{206} & $(2.6 \pm 0.4) \times 10^{-5}$ & $0.25 \pm 0.03$ \\ 
\isotope{Zn}{70}$\beta^-\beta^-$ & $(3.9 \pm 7.0) \times 10^{-5}$ & $< 0.1$ \\
\isotope{Zn}{64}$\beta^+\beta^+$ & $(1.3 \pm 3.9)\times 10^{-6}$ & $< 0.1$ \\
\isotope{Ge}{76}$\beta^-\beta^-$ & $(4.0 \pm 0.3) \times 10^{-8}$ & $< 0.1$ \\
\hline
\multicolumn{2}{|l|}{{\em Total Uncorrelated Rate}} & $14.4 \pm 4$ \\
\hline 
\multicolumn{3}{|l|}{{\em Correlated Flux Rate}} \\
\hline
\cns & & 4.85 \\
$\bar{\nu}_e e^{-}$-scattering & & 9.2$\times 10^{-5}$ \\
\hline
\multicolumn{2}{|l|}{{\em Total Correlated Rate}} & $4.85 \pm 0.25$ \\
\hline
\end{tabular}
\end{center}
\label{tab:backgrounds}
\end{table}%

\subsection{Neutrino-Induced Background}

As the extracted \cns~rate would be determined from its correlation with the reactor power, a rate-based analysis would not be able to discern between coherent and incoherent interaction channels.  The rate from $\bar{\nu}_e + e^-$ elastic scattering can be readily calculated, and is expected to yield $\sim 10^{-5}$ events/kg/day. Charged current interactions of $\bar{\nu}_e$ on zinc and germanium isotopes are limited because their thresholds are typically several MeV, and are expected to yield $\sim 0.004$ events/kg/day in Zn and $\sim 0.001$ events/kg/day in Ge.

\section{Sensitivity}

The strong suppression of the neutron background from the reactor core allows us to fit the data with with just two parameters: a reactor-correlated neutrino signal and a background that is uncorrelated with the reactor power output.  Although both the signal and background might posses subtle time structure due to the decay times of various isotopes, those should be sub-dominant to the correlation with the reactor power.  We perform two separate analyses; a {\em conservative} rate-only analysis, which looks at the integrated rate as a function of time, and a {\em best-case} analysis using both rate and shape, where the spectral information for the \cns~reaction is included in the signal extraction.  For a given signal rate $S$ and background $B$, the likelihood can be reconstructed from the combined probability of  seeing a number of counts in each bin $\rm{\bf N^{obs}}$. The number of observed events is reconstructed as a function of time and/or recoil energy, depending on the analysis.  The likelihood is reconstructed from simulated data and compared against either a background-only model or one which includes a signal correlated in time with the reactor thermal power. In both analyses, we make use of the Stan modeling language in order to assess the sensitivity of such a future measurement~\cite{stan}\footnote{We  use the morpho package as an interface to the Stan program(\href{https://github.com/project8/morpho})}.

\begin{table}[t]
\caption{Estimated sensitivity of the proposed detector to a CEvNS signal, for several possible background rates, run times, and reactor power conditions. Sensitivity is defined as the expected $1\sigma$ uncertainty on the signal rate divided by the signal rate. The input signal rate is assumed to be $\sim 5$ events per day at full reactor power.}
\begin{center}
\begin{tabular}{|c|c|c|c|c|c|}
\hline
Background & Run & Sensitivity & Sensitivity & Sensitivity & Sensitivity \\ 
& Time & (Rate Only, & (Rate Only,  & (Rate Only, & (Rate + \\
(events/day in ROI) & (years) & 80\% Full) & 60\% Full)  & 58\% Full, & Shape, \\
& &  & & 2\% Off) & 60\% Full) \\
\hline
15 & 1 & 22\% & 20\% & 17\% & 10\% \\
15 & 5 & 9\% & 7\% & 7\% & 5\% \\
\hline
36 & 1 & 31\% & 26\% & 22\% & 16\% \\
36 & 5 & 10\% & 10\% & 10\%  & 6\% \\
\hline
50 & 1 & 34\%  & 27\% & 25\% & 19\% \\
50 & 5 & 12\% & 11\% & 11\% & 8\% \\
 \hline
\end{tabular}
\end{center}
\label{tab:results}
\end{table}%

\begin{figure}[t]
\begin{center}
\includegraphics[width=0.89\textwidth]{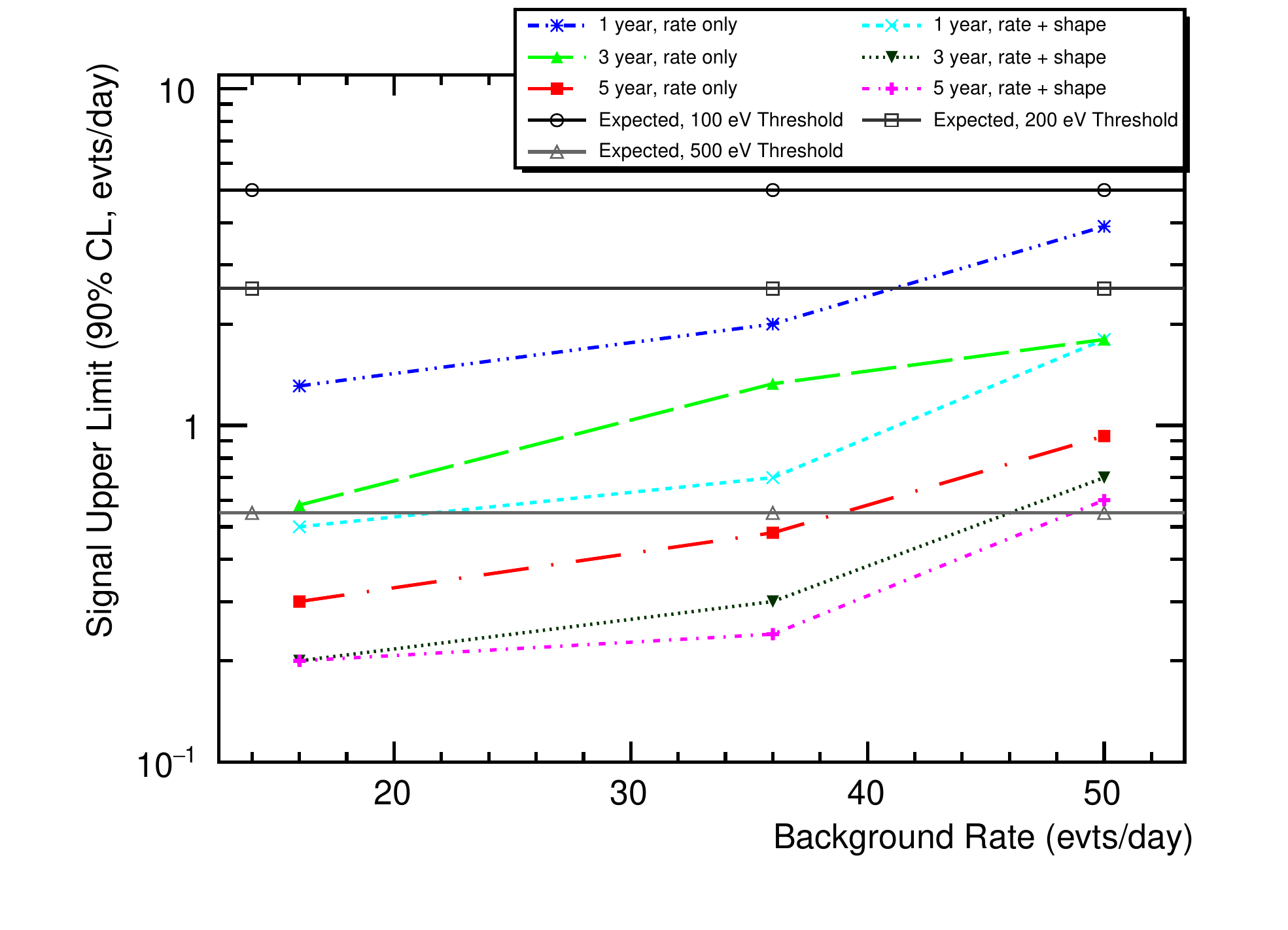}
\caption{The 90\% C.L. limit extracted in the case of zero input signal using a rate only (solid) and rate+shape (dashed) analyses. The reactor is modeled at full power 60\% of the time and half power for 40\%. The extracted limits are well below the expected 5 events/day expected from the Ricochet 10 kg target, assuming a 100 eV recoil energy threshold.}
\label{fig:90_pct_cl}
\end{center}
\end{figure}

When simulating data, we assume that the time structure of the background is flat, while that of the neutrino is tied solely to the power output from the two reactor cores.  Assuming, based on the average operating profile at the Chooz plant, that 40\% of the time one of the two reactors (B1 or B2, but not both) is off-line, we are able to extract the fraction of the signal that is correlated with the reactor flux~\cite{Abe:2012ar}.  We also explore the case where 20\% of the time one of the two reactors (B1 or B2, but not both) is off-line and the case when {\em both} reactors are offline for 2\% of the run cycle. In our analysis, we use a model that allows the background to have a component with exponential time dependence in addition to a flat component. We assume a total $\pm 5\%$ systematic uncertainty in the signal rate, which we attribute to uncertainties associated with maintaining a fixed energy threshold over time and daily variations in the thermal reactor flux, and thus is uncorrelated between different time bins in the analysis. This systematic is incorporated into our model by modifying the expected number of counts in each time bin by a gaussian envelope. The resulting likelihood function for the rate-only analysis is:

\begin{equation}
{\cal L}(S,B) = \sum_{i=0}^{N\_bins}{\cal P}
\{
{\rm \bf N^{obs}_i}|(X_i S(i) + B_1+B_2(i))\Delta t\}{\cal N} \{X_i|\mu,\sigma\}
\end{equation}

\noindent Where $B_1$ is the flat background, $B_2(i)$ is the background that decays exponentially in time, and $S(i)$ gives the signal in time bin i. Each time bin spans just over one week, and the time constant was assumed to be one year. Backgrounds are normalized such that $B_1+B_2(0)=B$, and $S(i)$is normalized such that the maximum value is $S$. The nuisance parameters $X_i$ are marginalized to obtain the likelihood. The parameters of the normal distribution ${\cal N}$ are set to $\mu=1$ and $\sigma=0.05$, in order to account for the $\pm 5\%$ systematic uncertainty. A similar likelihood function is used for the spectral shape analysis.

In the spectral shape analysis, all backgrounds except \isotope{H}{3} are assumed to be flat in our ROI when generating data. The analysis model then allows all backgrounds to have a component with exponential energy dependence. The energy resolution is assumed to be 15 eV, and is incorporated by computing the convolution product of the spectrum with a normal distribution. We use 100 eV-wide energy bins and the decay constant for the exponentially decaying background was assumed to be 10 keV.  Systematic errors are again incorporated into an uncorrelated $\pm 5\%$ uncertainty.

\begin{figure}[t]
\begin{center}
\includegraphics[width=0.89\textwidth]{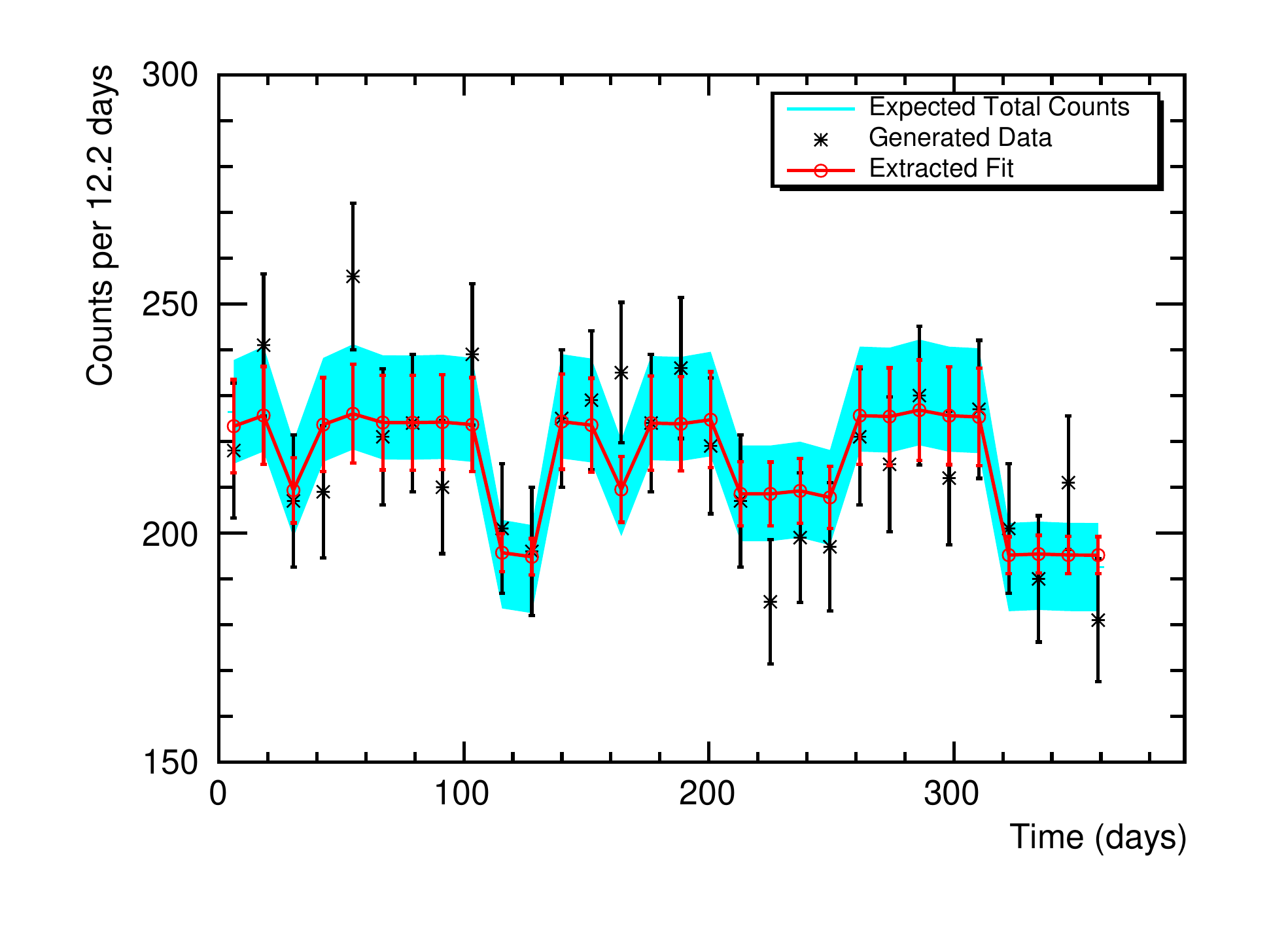}
\caption{The blue envelope is the expected number of counts for the true reactor power and background rate, with a 5\% variation allowed in the signal. The black and red points are respectively simulated data and the extracted fit. The background is 14.0 events/day and the signal is 5.0 evts/day. The reactor is modeled at full power 60\% of the time and half power for 40\%.}
\label{fig:time_data}
\end{center}
\end{figure}

Table~\ref{tab:results} shows the ability to extract the neutrino \cns~signal under different assumptions for the background level.  Although the baseline background levels considered here are based on those measured by the EDELWEISS detectors, the range considered is consistent with a number of different low threshold detectors, including CDEX~\cite{PhysRevD.93.092003} and CRESST~\cite{Strauss:2014aqw}.  Even under conditions of small signal-to-background ratios a strong neutrino signal can be extracted in a relatively small time scale (see Fig.~\ref{fig:time_data}).  By simulating the case where no signal is present, it is also possible to extract a lower limit on the signal a future Ricochet detector could detect. The 90\% confidence level of that lower signal limit is presented in Figure~\ref{fig:90_pct_cl}.

The Bayesian analysis described above has been cross-checked against the standard frequentist approach based on the profile likelihood ratio test statistics testing the null hypothesis (no \cns~events) against its alternative. Profiling over the background and signal efficiency systematics we were able to reproduce similar quantitative results as shown from the Bayesian analysis. Indeed, we found that we need at least 300 kg$\cdot$day of exposure to have a 95\% probability to reach a minimum discovery significance of 5 sigma with a 100 eVnr threshold, and that this required exposure decreases to less than 100 kg$\cdot$day if we reach a 10 eVnr energy threshold.

\section{Conclusion}

The combination of a luminous neutrino source, effective overburden and radiopure environment makes bolometric detection of the \cns~process extremely attractive at a location such as Chooz.  We have shown that even for a modest signal-to-noise ratio, a \cns~signal can be extracted with reasonable confidence, thanks to the leverage imparted from being able to cleanly distinguish neutrino signals from uncorrelated backgrounds.  We hope a demonstration of this technique will open the door to the myriad of scientific opportunities available from coherent neutrino scattering.

\section*{Acknowledgments}
We are grateful to the Heising-Simons Foundation for their generous support of the work presented here.  We wish to thank the MIT Electronics Research Society for access to their computing facilities.

\section*{References}
\bibliographystyle{iopart-num}
\bibliography{ChoozRicochet_revised}

\end{document}